\documentclass[aps,prl,amsmath,twocolumn,showpacs,superscriptaddress]{revtex4-1}
\usepackage{amsmath, amssymb, graphicx, bm}
\usepackage[usenames,dvipsnames,svgnames,table]{xcolor}
\usepackage[colorlinks=true,linkcolor=RoyalBlue,citecolor=RoyalBlue]{hyperref}

\begin{document}
\title{Spin Nernst Effect of Magnons in Collinear Antiferromagnets}

\author{Ran Cheng}
\affiliation{Department of Physics, Carnegie Mellon University, Pittsburgh, PA 15213}

\author{Satoshi Okamoto}
\affiliation{Materials Science and Technology Division, Oak Ridge National Laboratory, Oak Ridge, TN 37831}

\author{Di Xiao} 
\affiliation{Department of Physics, Carnegie Mellon University, Pittsburgh, PA 15213}


\begin{abstract}

In a collinear antiferromagnet with easy-axis anisotropy, symmetry guarantees that the spin wave modes are doubly degenerate. The two modes carry opposite spin angular momentum and exhibit opposite chirality. Using a honeycomb antiferromagnet in the presence of the Dzyaloshinskii-Moriya interaction, we show that a longitudinal temperature gradient can drive the two modes to opposite transverse directions, realizing a spin Nernst effect of magnons with vanishing thermal Hall current. We find that magnons around the $\Gamma$-point and the $\rm K$-point contribute oppositely to the transverse spin transport, and their competition leads to a sign change of the spin Nernst coefficient at finite temperature. Possible material candidates are discussed.

\end{abstract}

\maketitle

Recent years have seen a surge of interest in utilizing magnons for information encoding and processing~\cite{ref:Kajiwara,ref:Magn1,ref:Magn2,ref:BauerReview,ref:NL}.  Being an elementary excitation in magnetically ordered media, a magnon carries not only energy but also spin angular momentum~\cite{ref:Maekawa}.  The latter is of intrinsic interest in spintronics, since it would allow the transfer of spin information without Joule heating.  Such a realization has led to the emerging field of magnon spintronics~\cite{ref:Magn3}, in which magnons are expected to play similar roles as spin-$\frac{1}{2}$ electrons.  However, there is one caveat: while the electron spin forms an internal degree of freedom and is free to rotate, the magnon spin in a ferromagnet (FM) is fixed by its chirality, which can only be right-handed with respect to the magnetization.  

By contrast, it is well established that in a collinear antiferromagnet (AF) with easy-axis anisotropy, symmetry admits two degenerate magnon modes with opposite chirality~\cite{ref:AFMR}, and hence opposite spin~\cite{ref:Rezende,ref:SFZ}.  These two modes can be selectively excited and detected via both electrical~\cite{ref:Gomo,ref:STTAF,ref:SPAF} and optical~\cite{ref:Rasing,ref:Satoh,ref:VectorControl} means, which enables an internal space to encode binary information similar to the electron spin.  It is therefore possible to explore the magnonic counterparts of phenomena usually associated with the electron spin.  For example, a spin field-effect transistor of magnons using collinear AF has been recently proposed~\cite{ref:FET}, in which a rotation in the magnon spin space can be realized by a gate-tunable Dzyaloshinskii-Moriya interaction (DMI).

\begin{figure}[b]
	\includegraphics[width=\columnwidth]{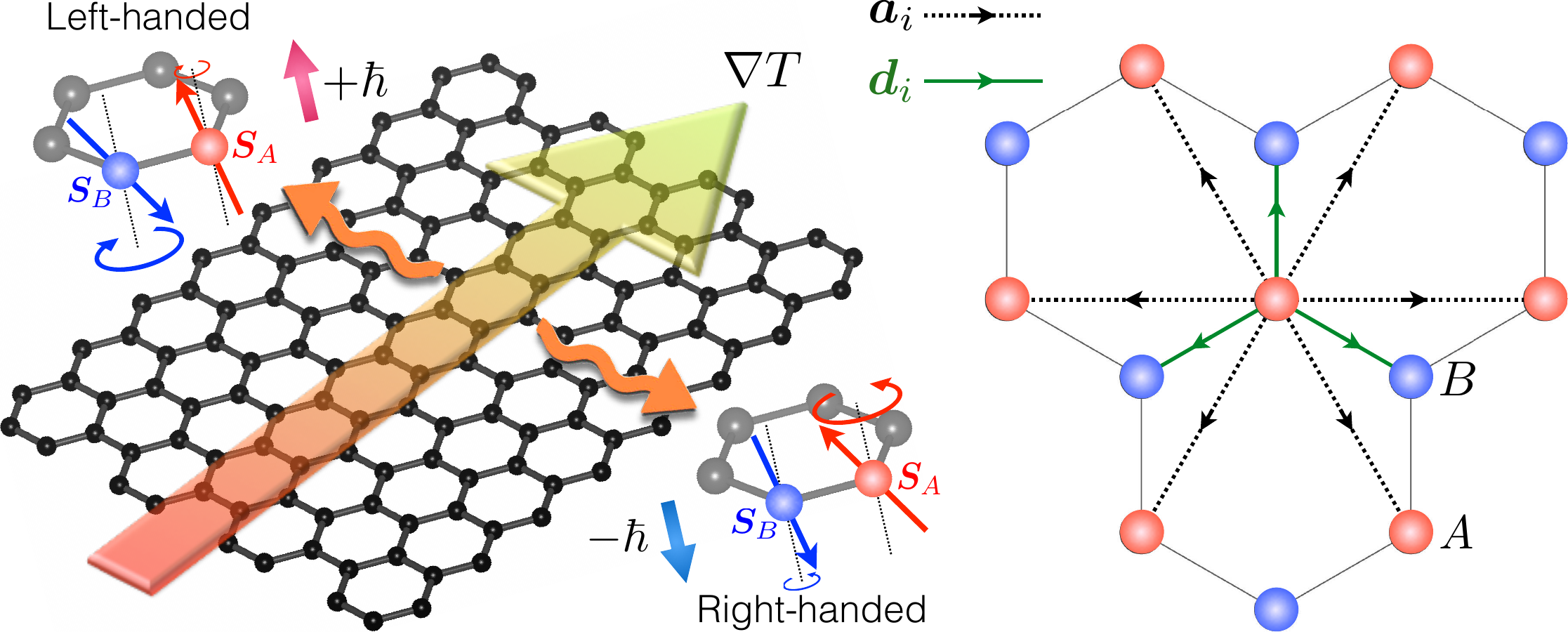}
	\caption{(Color online) Left: schematics of the magnon SNE. Right: the $J_1$--$D_2$ model on a honeycomb AF, the nearest neighbor and the second nearest-neighbor bonds are labeled by $\bm{d}_i$ and $\bm{a}_i$, respectively.}
	\label{fig:model}
\end{figure}

Drawing the above analogy, we theoretically demonstrate in this Letter a magnon spin Nernst effect (SNE) in a collinear AF, which is similar to the electron spin Hall effect~\cite{ref:SHE}.  The magnon SNE is intimately related to the magnon Hall effect~\cite{ref:ThermHE1,ref:ThermHE2,ref:RotMagWP,ref:Lifa,ref:ThermHE3,ref:ThermHE4}; it can be viewed as two opposite copies of the magnon Hall effect for each spin species, i.e., magnons with opposite spins flow in opposite transverse directions driven by an applied temperature gradient (Fig.~\ref{fig:model}).  We show that the SNE is realizable on a honeycomb lattice by including the second nearest-neighbor DMI.  The SNE coefficient is calculated through a semiclassical theory of magnon dynamics, supplemented by general symmetry analyses.  Finally, we propose MnPS$_3$~\cite{ref:MnPS3}, a layered magnetic compound, and its variances~\cite{ref:Nikhil} as possible material candidates to realize the magnon SNE.  Our results suggest that collinear AFs can serve as effective spin generators for both spin orientations in the same device, and provide a promising platform to explore novel caloritronic effects.

\emph{Model.}---Let us consider a collinear AF on a honeycomb lattice with the N\'{e}el order perpendicular to the hexagon plane, \textit{i.e.}, spins on the A and B sublattices satisfy $\bm{S}_A=-\bm{S}_B=S\hat{\bm{z}}$ in the ground state. Since the midpoint of the A--B link is an inversion center, the nearest neighbor DMI ($D_1$) vanishes~\cite{ref:DMI}.  However, the second nearest-neighbor DMI ($D_2$) is allowed by symmetry. The minimal spin Hamiltonian of such a system is
\begin{equation}
 H=J_1\sum_{\langle ij \rangle}\bm{S}_i\cdot\bm{S}_j+D_2\sum_{\langle\langle ij \rangle\rangle}\bm{\xi}_{ij}\cdot\bm{S}_i\times\bm{S}_j+\mathcal{K}\sum_iS_{iz}^2 \;, \label{eq:spinH}
\end{equation}
where $J_1>0$ is the nearest neighbor antiferromagnetic exchange coupling, $\mathcal{K}<0$ is the easy-axis anisotropy that ensures the N\'{e}el vector in the $\hat{\bm z}$ direction~\cite{ref:Ani}, and $\bm{\xi}_{ij} = 2\sqrt{3}\bm d_i \times \bm d_j = \pm\hat{\bm{z}}$ with $\bm d_i$ and $\bm d_j$ the vectors connecting site $i$ to its nearest neighbor site $j$ as shown in Fig.~\ref{fig:model}. We can include the second and the third nearest-neighbor exchange interactions $J_2$ and $J_3$ as well, but that will not alter the essential physics qualitatively. For simplicity, we have also set the length of the primitive vectors to be unity, $|\bm a_i| = 1$.

Using the Holstein-Primakoff transformation~\cite{ref:Nolting} and neglecting magnon-magnon interactions
\begin{subequations}
\begin{align}
 & S_{iA}^+\approx\sqrt{2S} a_i \;,\  S_{iA}^- \approx \sqrt{2S} a_i^\dag \;, \  S_{iA}^z=S-a_i^\dagger a_i \;,  \\
 & S_{iB}^+\approx\sqrt{2S} b_i^\dag \;,\  S_{iB}^- \approx \sqrt{2S} b_i \;, \  S_{iB}^z=b_i^\dag b_i - S \;,
\end{align}
\end{subequations}
we can express the spin Hamiltonian in the Nambu basis $\psi_k\equiv{a_k \brack b_k^\dagger}=\frac1{\sqrt{N}}\sum_{\bm{k}}e^{-i\bm{k}\cdot\bm{R}_i}{a_i \brack b_i^\dagger}$ as $H=\sum_k\psi^\dagger_k\mathcal{H}_k\psi_k$. Here, after discarding the zero-point energy, $\mathcal{H}_k$ reads
\begin{align}
 \mathcal{H}_k=S
 \begin{bmatrix}
  3J_1-K+D_2g(\bm{k}) & J_1f(\bm{k}) \\
  J_1f^*(\bm{k}) & 3J_1-K-D_2g(\bm{k})
 \end{bmatrix},
 \label{eq:Hamiltonian}
\end{align}
where $K=\mathcal{K}(2S-1)/S$, $f(\bm{k})=\sum\limits_{i}\exp(i\bm{k}\cdot\bm{d}_i)$ and $g(\bm{k})=\sum\limits_{i\in\mathrm{odd}}2\sin(\bm{k}\cdot\bm{a}_i)$ with $\bm{a}_i$ the vectors linking second nearest neighbors (see Fig.~\ref{fig:model}).  We note that $g(\bm k)$ is an odd function of $\bm k$.

To diagonalize Eq.~\eqref{eq:Hamiltonian}, we perform a Bogoliubov transformation $\alpha_k=u_ka_k-v_kb_k^\dagger$ and $\beta_k=u_kb_k-v_ka_k^\dagger$ that mixes magnons on different sublattices~\cite{ref:Nolting}. The Heisenberg equation of motion (EOM) $i\hbar\dot{\alpha}_k=[\alpha_k, \mathcal{H}_k]$ yields the eigenequations of the Bogoliubov wave function $\Psi^\alpha={u_k \brack v_k}$ of the $\alpha$ mode as
\begin{align}
 \hbar\sigma_z\omega_{\alpha}\Psi^\alpha=\left( a\mathrm{I}+b\sigma_x+c\sigma_y+d\sigma_z \right)\Psi^\alpha, \label{eq:Eigen}
\end{align}
where $a=S(3J_1-K)$, $b=SJ_1\mathrm{Re}f(\bm{k})$, $c=SJ_1\mathrm{Im}f(\bm{k})$, and $d=SD_2g(\bm{k})$. Equation~\eqref{eq:Eigen} is akin to a Schr\"{o}dinger equation except the $\sigma_z$ factor on its left hand side, which is ascribed to the bosonic commutation relation $[\alpha_k,\alpha_k^\dagger]=\delta_{kk'}$. This feature enables a hyperbolic parametrization of Eq.~\eqref{eq:Eigen}: $a=\ell\cosh\theta$, $b=\ell\sinh\theta\cos\phi$, $c=\ell\sinh\theta\sin\phi$. The spectrum is then $\hbar\omega_\alpha=d\pm\ell$, and the corresponding eigenvectors are
\begin{equation}
\Psi_+^\alpha=
\binom{\cosh\frac{\theta}{2}}{-\sinh\frac{\theta}{2}e^{i\phi}},\quad
\Psi_-^\alpha=
\binom{-\sinh\frac{\theta}{2}}{\cosh\frac{\theta}{2}e^{i\phi}},
\label{eq:eigenvectors}
\end{equation}
which respects the generalized orthonormal conditions $\langle\Psi_{\pm}^\alpha|\sigma_z|\Psi_{\pm}^\alpha\rangle=\pm1$ and $\langle\Psi_{\pm}^\alpha|\sigma_z|\Psi_{\mp}^\alpha\rangle=0$.  Since we are interested in quasiparticle excitations, we will keep the positive branch and drop the negative one.  In the same manner, the Heisenberg EOM $i\hbar\dot{\beta}_k=[\beta_k, \mathcal{H}_k]$ yields a similar eigenequation, but with the $\sigma_z$ term on the right hand side of Eq.~\eqref{eq:Eigen} flipping sign.  Nonetheless, the associated eigenvectors are exactly the same as Eq.~\eqref{eq:eigenvectors}, since neither $\theta$ nor $\phi$ depend on $D_2$.  Together, the energy spectrum of the two magnon branches are given by
\begin{equation}
 \hbar\omega_{\alpha,\beta}= S\Bigl[\sqrt{(3J_1-K)^2 - J_1^2|f(\bm k)|^2} \pm D_2g(\bm k)\Bigr] \;,
 \label{eq:spectrum}
\end{equation}
where the plus (minus) sign corresponds to the $\alpha$ mode ($\beta$ mode). While the $D_2$ term breaks the degeneracy, it does not change the wave functions. Note that for sufficiently large $D_2$ (comparable to $J_1$), our theory breaks down as the ground state is no longer a collinear AF but a spin spiral. Throughout this Letter, we will restrict to the regime where the collinear order is preserved.

The physical meaning of the two magnon modes can be intuitively understood using the semiclassical picture described by the Landau-Lifshitz equation~\cite{ref:AFMR}. By identifying $S_i^+$ and $S_i^-$ as generating opposite precessions on site $i$, we see that both $\bm{S}_A$ and $\bm{S}_B$ precess in the right-handed (left-handed) way in the $\alpha$ mode ($\beta$ mode), as illustrated in Fig.~\ref{fig:model}. Consequently, the two modes can be distinguished by their opposite chirality. In the semiclassical picture, it is also clear why the negative branches are redundant: $S_i^+e^{i\omega t}$ and $S_i^-e^{-i\omega t}$ describe the same spin precession since $\mathrm{Re}[S_i^+e^{i\omega t}]=\mathrm{Re}[S_i^-e^{-i\omega t}]$  with $S_i^{\pm}=(S_i^x\pm iS_i^y)/2$. Moreover, since $u_k=\cosh\theta/2$ and $v_k=-e^{i\phi}\sinh\theta/2$ switch roles between $\alpha_k$ and $\beta_k^\dagger$, the ratio of sublattice magnon densities $\langle a_i^\dagger a_i\rangle/\langle b_i^\dagger b_i\rangle$ in the two modes are reciprocal to each other, which, as schematically shown in Fig.~\ref{fig:model}, corresponds to different precessional cone angles of $\bm{S}_A$ and $\bm{S}_B$.

The magnon chirality is intimately related to its spin.  Since the $J_1$-$D_2$ model preserves the rotational symmetry around the $z$-axis, the $z$-component of the total spin $S^z=\sum_i(S_{iA}^z+S_{iB}^z)$ should be conserved. By inserting the Holstein-Primakoff transformation into $S^z$, we obtain $S^z=\sum_kS^z_k=\sum_k(-a_k^\dagger a_k+b_k^\dagger b_k)$.  Since $S^z_k$ is diagonal in the Nambu basis, it commutes with the Hamiltonian $[S^z_k, H]=0$ where $H=\sum_k\psi^\dagger_k\mathcal{H}_k\psi_k$.  By invoking the Bogoliubov transformation, we further obtain
\begin{align}
	S^z=\sum\nolimits_k(-\alpha_k^\dagger\alpha_k+\beta_k^\dagger\beta_k), \label{eq:Sz}
\end{align}
thus $\langle0|\alpha_kS^z\alpha_k^\dagger|0\rangle=-1$ and $\langle0|\beta_kS^z\beta_k^\dagger|0\rangle=+1$ with $|0\rangle$ denoting the magnon vacuum.  This indicates that a quantum of the $\alpha$ magnon ($\beta$ magnon) carriers $-1$ (+1) spin angular momentum along the $\hat{\bm{z}}$-direction, i.e., the spin-$z$ component is \textit{locked} to the magnon chirality and is independent of the momentum $\bm{k}$.  We note that this relation is specific to the symmetry of our model.  For example, an in-plane easy axis anisotropy destroys the rotational symmetry around the $z$-axis, and will spoil this relation.  

\emph{AF magnon dynamics.}---Since the two magnon modes are completely decoupled, we can treat the dynamics of each independently so long as the $\sigma_z$ factor in Eq.~\eqref{eq:Eigen} is properly taken care of. Let us consider a magnon wave packet in the positive branch $|W\rangle=\int d\bm{k}\,w(\bm{k},t)|\Psi_+(\bm{k})\rangle$ localized around the center $(\bm{r}_c,\bm{k}_c)$ in the phase space, where $\bm{r}_c=\langle W|\bm{r}|W\rangle$ and $\bm{k}_c=\int d\bm{k}|w(\bm{k})|^2\bm{k}$. The definition of $|W\rangle$ does not specify whether it represents a spin-up or a spin-down magnon because the two modes have the same wave function.  The magnon dynamics can be obtained by taking the variational derivative of the functional Lagrangian $\mathcal{L}=\langle W|i\hbar\sigma_z\frac{d}{dt}|W\rangle-\langle W|\mathcal H^*|W\rangle$ with respect to $\bm{r}_c$ and $\bm{k}_c$~\cite{ref:RotMagWP,ref:Review}. In particular, the EOM of $\bm r_c$ is given by
\begin{align}
 \dot{\bm{r}}_c=\frac{\partial\omega(\bm{k}_c)}{\partial\bm{k}_c}+\frac1\hbar\bm{\nabla}U(\bm{r}_c)\times\bm{\Omega}(\bm{k}_c), \label{eq:EOM}
\end{align}
where $U(\bm{r})$ is the potential felt by the magnons, and $\Omega(\bm{k})$ is the Berry curvature
\begin{align}
 \bm{\Omega}(\bm{k})&=-\mathrm{Im}\langle\bm{\nabla}\Psi_+(\bm{k})|\times\sigma_z|\bm{\nabla}\Psi_+(\bm{k})\rangle \notag\\
 &=\frac12\sinh\theta(\bm{\nabla}\theta\times\bm{\nabla}\phi),
\end{align}
which only has an out-of-plane component $\bm{\Omega}(\bm{k})=\Omega(\bm{k})\hat{\bm{z}}$ in two dimensions. It is the Berry curvature that gives rise to a transverse motion of the magnon wave packet and leads to a Hall response.

Before turning to any specific transport effect, let us explore the symmetry properties of our $J_1$-$D_2$ model and find out what ensures a transverse transport.  Given the N\'{e}el ground state, we expand the spin Hamiltonian~\eqref{eq:spinH} to the quadratic order in $\delta\bm{S}_{A}=\bm{S}_{A}-\hat{\bm{z}}$ and $\delta\bm{S}_{B}=\bm{S}_{B}+\hat{\bm{z}}$ as $H=H_J+H_K+H_D$, where (set $S=1$)
 \begin{align}
 H_J &= \sum_{\langle AB \rangle} J_1(1-\delta S_A^z+\delta S_B^z+\delta\bm{S}_A\cdot\delta\bm{S}_B) \;, \notag \\
 H_K &= \sum_{A,B} 2\mathcal{K}(1+\delta S_A^z-\delta S_B^z)+\mathcal{K}[(\delta S_A^z)^2+(\delta S_B^z)^2] \;, \notag \\
 H_D &= \sum_{\langle\langle AA' \rangle\rangle} D_2(\delta S_A^x\delta S_{A'}^y-\delta S_A^y\delta S_{A'}^x)-(A\rightarrow B) \;, \notag
 \end{align}
with $\langle AB \rangle$ denoting nearest neighbor sites and $\langle\langle AA' \rangle\rangle$ second nearest-neighbor sites.  Since we are interested in the symmetry properties of magnons, all symmetry operations act only on the magnon parts $\delta\bm{S}_{A}$ and $\delta\bm{S}_{B}$ while leaving the N\'{e}el ground state unchanged.  

We first analyze the symmetry properties in the absence of the DMI.  It is easy to see that $H_J+H_K$ is invariant under the combined symmetry of time-reversal ($\mathcal T$) and a $180^\circ$ rotation around the $\hat{\bm{x}}$-axis in the spin space ($c_x$).  By demanding the EOM invariant under $\mathcal Tc_x$, we obtain $\omega(\bm k) = \omega(-\bm k)$ and $\bm{\Omega}(\bm{k})=-\bm{\Omega}(-\bm{k})$. On the other hand, $H_J+H_K$ breaks the inversion symmetry (not true for a ferromagnet); hence a nonzero Berry curvature can develop even without the DMI~\cite{footnote}.

The $H_D$ term apparently breaks the $Tc_x$ symmetry. However, as mentioned earlier, the wave functions are independent of $D_2$, thus the Berry curvature is not affected by $D_2$.  What $H_D$ really does is invalidating the relation $\omega(\bm{k}) = \omega(-\bm{k})$ as can be seen from Eq.~\eqref{eq:spectrum}.  This will cause a population imbalance between $\bm k$ and $-\bm k$ states, leading to a net Berry curvature and hence a transverse current for each spin species.  Since the $D_2$ correction to $\omega(\bm k)$ is opposite for the two modes, the transverse thermal current should vanish identically. Therefore, the net effect should be a spin-Hall like phenomenon.

It is useful to compare the role of the DMI in a honeycomb AF with its FM counterpart~\cite{ref:Owerre,ref:Kim}.  In a honeycomb FM, both the $Tc_x$ and the inversion symmetries are kept by $H_J+H_K$ so that the Berry curvature is identically zero before turning on the DMI.  The $D_2$ term breaks $Tc_x$ and changes the band topology, which opens a finite gap at the Dirac points and hence a nonzero Berry curvature, giving rise to a magnon Hall effect~\cite{ref:Owerre,ref:Kim}; the physics parallels exactly Haldane's quantum anomalous Hall model~\cite{ref:Haldane}. By contrast, the gap opening in our honeycomb AF occurs at the $\Gamma$-point because of the easy-axis anisotropy $K$, whereas the DMI does not affect the band topology. 

\emph{Spin Nernst effect.}---Magnons are charge neutral, so they cannot be driven by an electric field. Nevertheless, by introducing an in-plane temperature gradient $\bm{\nabla}T$, one can create a longitudinal magnon current.  Because of the Berry curvature, a magnon Hall current is induced for each individual spin species~\cite{ref:RotMagWP,ref:DX} as
\begin{align}
 \bm{j}_\lambda=\frac{k_B}{\hbar}\hat{\bm{z}}\times\bm{\nabla}T & \int [d\bm{k}]\Omega(\bm{k})\left\{\rho_{\lambda}(\bm{k})\ln\rho_{\lambda}(\bm{k})\right. \notag\\
 &\quad\left.-[1+\rho_{\lambda}(\bm{k})]\ln[1+\rho_{\lambda}(\bm{k})] \right\}, \label{eq:MHcurrent}
\end{align}
where $\lambda=\downarrow$ ($\uparrow$) refers to the $\alpha$ mode ($\beta$ mode), $[d\bm{k}]=d^2k/(2\pi)^2$, $k_B$ is the Boltzmann constant, and $\rho_\lambda=1/(e^{\hbar\omega_\lambda/k_BT}-1)$ is the Bose-Einstein distribution function with the chemical potential taken to be zero (since the magnon number is not conservative). As can be anticipated from the symmetry argument shown earlier, $\bm{j}_{\uparrow}=\bm{j}_{\downarrow}=0$ if $D_2$ vanishes. This is because when $D_2=0$, $\omega_\uparrow(\bm{k})=\omega_\downarrow(\bm{k})=\omega_0(\bm{k})$ is even, so is $\rho_\lambda(\bm{k})$; but $\Omega(\bm{k})$ is odd; thus the integration of Eq.~\eqref{eq:MHcurrent} vanishes. A finite $D_2$ leads to an opposite change of the spectrum $\hbar\omega_{\uparrow/\downarrow}(\bm{k})=\hbar\omega_0(\bm{k})\mp SD_2g(\bm k)$ with $g(\bm k)=-g(-\bm k)$, whereas the Berry curvature remains unchanged.

In the linear response regime, the SNE current can be written as $\bm{j}_{_{SN}}=\hbar(\bm{j}_\uparrow-\bm{j}_\downarrow)\equiv\alpha_{xy}^s\hat{\bm{z}}\times\bm{\nabla}T$, where $\alpha_{xy}^s$ is the SNE coefficient. In general, an analytic expression of $\alpha_{xy}^s=\alpha_{xy}^s(D_2,T)$ is not available. Nevertheless, we can derive an approximate expression of $\alpha_{xy}^s$ in the limit of $D_2\ll J_1$. Expanding $\rho_\lambda$ to linear order in $D_2$, we obtain from Eq.~\eqref{eq:MHcurrent} that
\begin{align}
	\alpha_{xy}^s\approx\frac{2\hbar D_2}{k_BT^2}\int d\omega\frac{\omega e^{\hbar\omega/k_BT}}{(e^{\hbar\omega/k_BT}-1)^2}\mathcal{D}(\omega)\Omega(\omega), \label{eq:alpha}
\end{align}
where $\mathcal{D}(\omega)=\int_{BZ} [d\bm{k}]\delta[\omega-\omega(\bm{k})]$ is the density of states (DOS) and $\Omega(\omega)=\int_{BZ} [d\bm{k}]\delta[\omega-\omega(\bm{k})]\Omega(\bm{k})$ is the density of the Berry curvature.

\begin{figure}[t]
	\centering
	\includegraphics[width=\columnwidth]{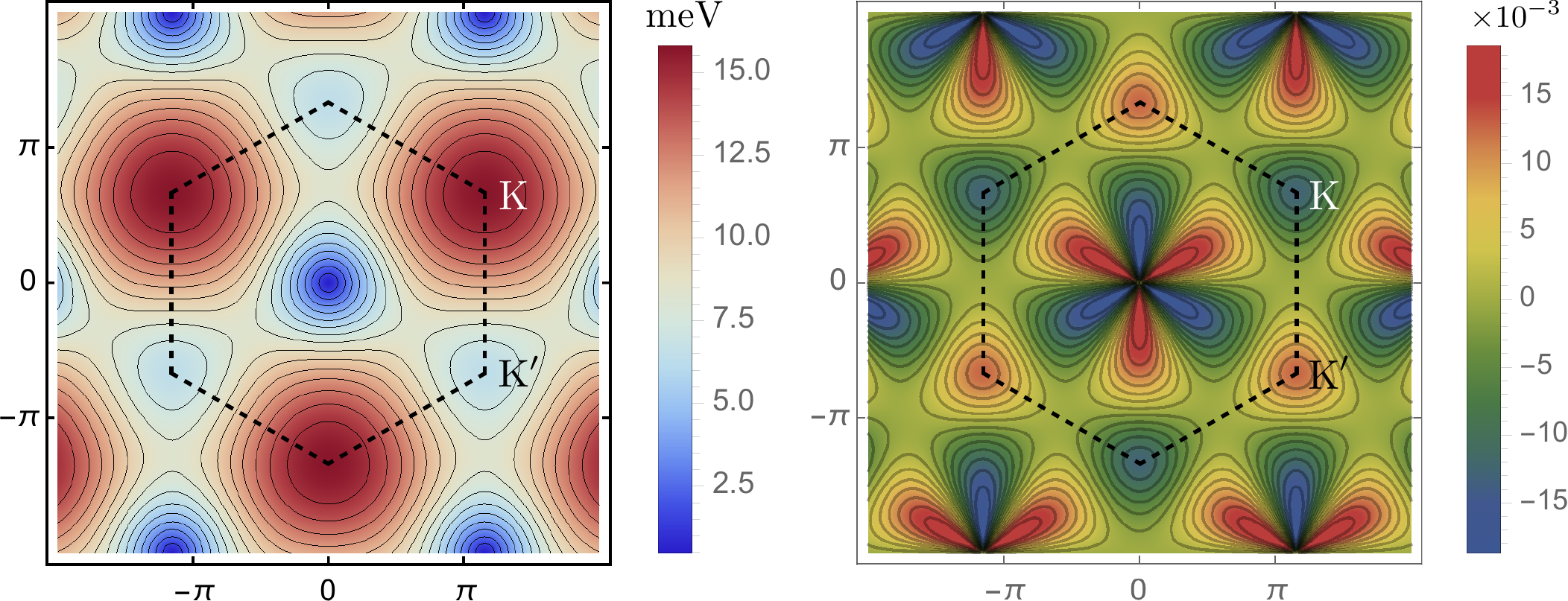}
	\caption{(Color online) Dispersion and Berry curvature of the spin-down (right-handed) magnon mode with $J_1=1.54$, $J_2=0.14$, $J_3=0.36$, and $KS=-0.0086$ taken from MnPS$_3$~\cite{ref:J}, assuming $D_2=0.36$. Numbers are in units of meV.}
	\label{fig:band}
\end{figure}

\emph{Material realization.}---Our theoretical proposal of the SNE could be experimentally tested in a number of honeycomb mangets.  One possibility is Mn-based trichalcogenide, such as MnPS$_3$~\cite{ref:MnPS3}.
In this compound, the magnetic moments of Mn ions are arranged on layered honeycomb lattices, and are coupled antiferromagnetically.  In addition, the Mn ions are half-filled with the high spin state $S=\frac{5}{2}$, so the quantum fluctuation in these materials is not as important as that in spin-$\frac{1}{2}$ systems.  
It has been well established that to properly capture the magnon dynamics in such systems, $J_1$ is not enough; one needs to include the second and the third nearest-neighbor exchange couplings $J_2$ and $J_3$ as well~\cite{ref:MnPS3,ref:Nikhil}. Nonetheless, $J_2$ and $J_3$ do not invalidate the symmetries of the spectrum and the Berry curvature, they only entail quantitative changes.  

In the following we will treat $D_2$ as a tuning parameter in our calculation since its actual value is not available in existing literature.  Figure~\ref{fig:band} shows the spectrum of the spin-down magnon (\textit{i.e.}, the $\alpha$ mode) and the associated Berry curvature using material parameters adapted from MnPS$_3$~\cite{ref:MnPS3,ref:J}, assuming $D_2=0.36$ meV.  Note that this $D_2$ is well below the critical value for the spin texture formation so that the N\'{e}el ground state is protected. The odd parity shown in Fig.~\ref{fig:band} is consistent with our symmetry analysis.

\begin{figure}[t]
	\centering
	\includegraphics[width=0.96\columnwidth]{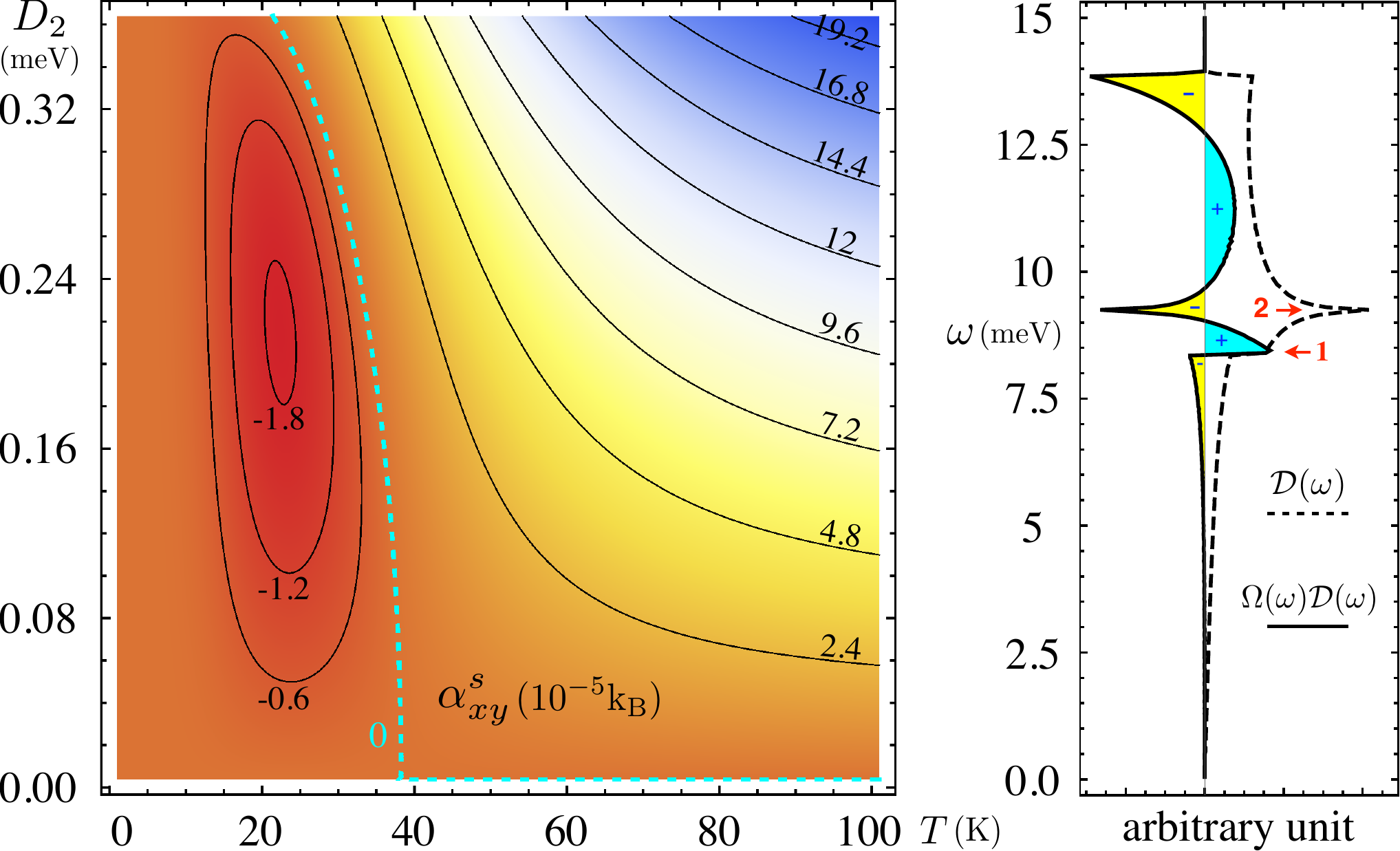}
	\caption{(Color online) Left: the SNE coefficient as a function of temperature and $D_2$ based on materials parameters of MnPS$_3$~\cite{ref:MnPS3,ref:J}. Right: the DOS $\mathcal{D}(\omega)$ and the joint density $\Omega(\omega)\mathcal{D}(\omega)$ for the spin-up mode at $D_2=0.21$ meV, where the two von Hove singularities are marked in red. For the spin-down mode, $\mathcal{D}(\omega)$ is the same but $\Omega(\omega)$ switches sign.}
	\label{fig:phase}
\end{figure}

Figure~\ref{fig:phase} shows the numerical result of $\alpha_{xy}^s$ of MnPS$_3$ as a function of temperature and $D_2$ using Eq.~\eqref{eq:MHcurrent}. Note that our model analysis  based on the linear spin wave theory is only valid in the temperature range much lower than the N\'{e}el temperature, which is estimated to be 160$\sim$230 K~\cite{ref:Nikhil}. A striking feature is that the SNE coefficient $\alpha_{xy}^s$ is not monotonic in either $D_2$ or $T$.  For fixed $D_2$, $\alpha_{xy}^s$ first goes negative and then bends up, and finally experiences a sign reversal with an increasing temperature. Such a pattern persists throughout the range of $D_2$ we explored, and a maximum negative value of $\alpha_{xy}^s$ takes place around $T=23$ K and $D_2=0.21$ meV.

The sign change of $\alpha_{xy}^s$ can be qualitatively understood with the help of Eq.~\eqref{eq:alpha}.  We plot the DOS and the joint density $\Omega(\omega)\mathcal{D}(\omega)$ in Fig.~\ref{fig:phase} for $D_2=0.21$ meV. For the spin-up mode (spin-down mode), the $\rm K'$-point ($\rm K$-point) of the spectrum is a local minimum, and the midpoint between $\rm K'$ ($\rm K$) and $\Gamma$ is a saddle point. These features give rise to two von Hove singularities in the DOS. We see that the Berry curvature flips sign across the von Hove singularities, which indicates that magnons around the $\Gamma$-point contribute to $\alpha_{xy}^s$ oppositely comparing to magnons from the $\rm K'$ valley (or $\rm K$ valley, whichever forms a local minimum in the spectrum depending on the spin of the mode). Raising temperature increases the relative contribution of the latter, which competes with the former and eventually leads to a sign change of $\alpha_{xy}^s$.

Finally, we notice that besides the simple N\'{e}el state, a variety of ordered ground states, including both FM and AF zigzag configurations, have been observed in transition-metal trichalcogenides~\cite{ref:Brec}. In the presence of DMIs, this family of compounds might exhibit rich thermomagnetic behavior, rendering them an ideal playground for chiral magnon transport. 

In summary, we have theoretically demonstrated a magnon spin Nernst effect in a collinear honeycomb antiferromagnet with out-of-plane N\'{e}el ground state, and have proposed monolayer MnPS$_3$ as a possible candidate to realize this effect.  The underlying physics is attributed to the breaking of the $\mathcal{T}c_x$ symmetry by the second nearest-neighbor DMI, which changes the parity of the spectrum but does not affect the Berry curvature. 
The ability to generate a pure transverse spin current devoid of a thermal current would be of great interest in magnon spintronics.

\textit{Note added.}---After completion of the bulk of this work (see, e.g., the brief announcement in Ref.~\cite{ref:March}), a related work has appeared, in which a spin Nernst effect of spinons is discussed~\cite{ref:Kim}. However, they considered a honeycomb ferromagnet where the SNE is only possible in the disordered phase, whereas in our case the SNE is found in the ordered AF phase. The governing physics is of completely different regimes.

\begin{acknowledgments}
	We are grateful to Ying Ran for insightful discussions. We would also like to thank Igor Barsukov, Matthew W. Daniels, Nikhil Sivadas, and Jimmy Zhu for useful comments. R.C.\ and D.X.\ were supported by the Department of Energy, Basic Energy Sciences, Grant No.~DE-SC0012509.  S.O.~acknowledges support by the U.S. Department of Energy, Office of Science, Basic Energy Sciences, Materials Sciences and Engineering Division.
\end{acknowledgments}


\begin{thebibliography}{20}
\bibitem{ref:Ani}It is known that the single-ion anisotropy in the form of Eq.~\eqref{eq:spinH} vanishes for $S=\frac{1}{2}$.  However, a collinear AF ordering with a finite magnon gap could be stabilized by an Ising-type anisotropy in the exchange coupling. This term does not change the symmetry of the Hamiltonian and, consequently, the existence of the magnon spin Nernst effect.
\bibitem{footnote} If the system also has inversion symmetry, then one would have $\bm{\Omega}(\bm{k})=-\bm{\Omega}(-\bm{k})$. This, together with the $\mathcal Tc_x$ symmetry, renders the Berry curvature zero.
\bibitem{ref:J} The value of $J$'s are taken from Ref.~\cite{ref:MnPS3}.  Note that there is a factor of 2 difference due to the different conventions used in the Heisenberg Hamiltonian. The single-ion anisotropy term is also converted accordingly.
\end{thebibliography}
\end{document}